\documentclass
[superscriptaddress,secnumarabic,amssymb,amsmath,nobibnotes,aps,prd,nofootinbib,onecolumn]{revtex4}%
\usepackage{graphicx}
\usepackage{epstopdf}
\usepackage{epsf}
\usepackage{bm}
\usepackage{amsmath}
\usepackage{amsfonts}
\usepackage{amssymb}
\usepackage{physics}

\providecommand{\U}[1]{\protect\rule{.1in}{.1in}}

\newcommand{\be}{\begin{equation}}
\newcommand{\ee}{\end{equation}}

\newcommand{\mincir}{\raise
-3.truept\hbox{\rlap{\hbox{$\sim$}}\raise4.truept\hbox{$<$}\ }}
\newcommand{\magcir}{\raise
-3.truept\hbox{\rlap{\hbox{$\sim$}}\raise4.truept\hbox{$>$}\ }}

\newcommand{\ud}{\,\mathrm{d}}

\usepackage{comment}
\newcommand{\sgn}{\mathop{\mathrm{sgn}}}
\newcommand{\atanh}{\mathop{\mathrm{tanh^{-1}}}}
\usepackage[linktocpage=true]{hyperref}
\hypersetup{colorlinks=true,citecolor=blue,linkcolor=blue,urlcolor=blue}

\begin{document}

\title{General solutions to $\mathcal{N}$-field cosmology with exponential potentials
}

\author{Perseas Christodoulidis}
\email{p.christodoulidis@rug.nl}
\affiliation{Van Swinderen Institute for Particle Physics and Gravity, University of
\ Groningen, Nijenborgh 4, 9747 AG Groningen, The Netherlands}

\begin{abstract}
We construct the general analytical solution for the $\mathcal{N}$-field product-exponential potential in an expanding FLRW background. We demonstrate the relevance of this analytical solution in more general contexts for the derivation of estimates for the transitional time between an arbitrary initial state and the slow-roll solutions. In certain cases, these estimates can also be used to demonstrate the non-linear convergence towards the slow-roll solutions. In addition, we extend the solution to include stiff matter as well.
\end{abstract}

\maketitle 

\tableofcontents

\section{Introduction}

Inflation is the leading paradigm for the origin of structure formation in an initially isotropic and homogeneous universe \cite{Akrami:2018odb}. It was originally proposed as an attempt to evade the fine-tunings of the standard cosmological model by theorizing a period of accelerating expansion in the very early universe \cite{Starobinsky:1980te,Sato:1981ds,Sato:1980yn,Kazanas:1980tx,Guth:1980zm,Linde:1981mu,Albrecht:1982wi}. Despite the simplicity of this idea, the non-linear nature of scalar-field equations 
makes the quest for analytical solutions impossible; without further simplifications we have to rely entirely on numerical tools. Although efficient numerical codes have been developed during the last decade, for instance the publicly available codes \cite{Price:2014xpa,Dias:2015rca,Dias:2016rjq,Ronayne:2017qzn,Mulryne:2016mzv}, (approximate) analytical solutions are desirable for a number of reasons: if inflation solves the fine-tuning problems by utilizing an attractor mechanism, then this should be shown analytically rather than numerically in a case-by-case analysis. In addition, for multiple fields calculation of non-Gaussianities and other higher order correlators is computationally demanding and therefore simple analytical expressions may provide more insight on how to better search for viable models. 

The only single-field examples in the literature that admit general analytical expressions are the exponential \cite{Salopek:1990jq,Chimento:1998ju,Andrianov:2011fg,Elizalde:2004mq,deRitis:1990ba,Russo:2004ym} and trigonometric hyperbolic potentials \cite{Bertacca:2007ux,Piedipalumbo:2011bj}. It has been shown in Refs.~\cite{Basilakos:2011rx,Paliathanasis:2014zxa} that these potentials respect the symmetries of the mini-superspace metric (spanned by the scale factor and the field) leading to an integrable system \cite{Tsamparlis:2011wg,Tsamparlis:2011wf}. Similarly, general multi-field solutions have been constructed for two non-interacting scalar fields in a flat field-space, where one is massless and the other has an exponential potential \cite{Chimento:1998ju,Zhang:2009mm}, and for hyperbolic trigonometric potentials in a hyperbolic field-space manifold \cite{Paliathanasis:2014yfa,Anguelova:2018vyr,Paliathanasis:2018vru}. Specifically, in Refs.~\cite{Paliathanasis:2014yfa,Tsamparlis:2011nsv} it was shown that completely integrable two-field systems require a field-manifold of constant curvature, hence the flat and hyperbolic cases. Apart from the previous solutions, more have been derived, e.g.~late-time scaling solutions\footnote{For scaling solutions the first slow-roll parameter $\epsilon \equiv -\dot{H}/H^2$ is constant and the scale factor increases in a power-law fashion $a(t)\sim t^{b}$. They belong to the class of power-law inflation \cite{Lucchin:1984yf}
.} \cite{Halliwell:1986ja,Steinhardt:1999nw,Nunes:2000yc,Liddle:1998jc,Malik:1998gy,Copeland:1999cs} and particular exact solutions using the superpotential method \cite{Easther:1993qg,Chervon:1997yz,Chervon:2017kgn,Martin:2013tda,Arefeva:2009tkq,Bazeia:2013uba,Dimakis:2019qfs,Moraes:2014cha}.


The aim of this work is the derivation of general solutions for an arbitrary number of fields. These solutions can be used to derive estimates for the transition time (or field displacement) from arbitrary initial states to inflation, or alternatively they can be used to demonstrate non-linear convergence towards the slow-roll solutions. 

The paper is organized as follows: in Sec.~\ref{sec:gs} we derive the general solution for a product-exponential potential. In Sec.~\ref{subsec:ab} we provide analytical estimates for the duration of the passage from an arbitrary initial state towards inflation. 
In Sec.~\ref{sec:generalization} we continue the investigation of general solutions in the presence of a fluid. 
Finally, in Sec.~\ref{sec:concl} we summarize our main results.

\textbf{Conventions:} dots ($\dot{}$) and primes ($'$) indicate derivatives with respect to the cosmic time and number of e-folds respectively; Greek letters refer to spacetime indices while Latin ones to field-space indices; the spacetime metric has the mostly plus signature $(-,+,+,+)$; we assume the Einstein's convention for repeated indices; $N$ is the e-folding number and $\mathcal{N}$ denotes the number of fields.

\section{The general analytical solution} \label{sec:gs}

\subsection{Earlier works}

Simple models of the early and late universe assume an action with minimal derivative couplings and canonical kinetic terms 
\begin{equation}
S = \int \ud^4 x \sqrt{-g} \left(  \frac{\kappa}{2} R - \frac{1}{2} \delta_{ij} \partial_{\mu}  \phi^i \partial^{\mu} \phi^j -V \right)  \, ,
\end{equation}
where the Latin indices label the number of fields, running from 1 to $\mathcal{N}$, and $\kappa=8\pi G/c^2$ (we will set $\kappa=1$ from now on). Variation of the action with an FLRW ansatz for the metric
\begin{equation}
\ud s^2 = -\ud t^2 + a(t)^2  \ud \vec x \cdot \ud \vec x \, ,   
\end{equation}
yields the two linearly independent Einstein's equations for the unknown metric function $a(t)$ (one evolution equation and one constraint) and the generalized Klein-Gordon equations for the scalar fields $\phi_i(t)$. This system of equations can be be written in compact form with the definition of the Hubble parameter $H= \ud (\ln a) / \ud t$ 
\begin{align} 
	\label{eq:fc} &3H^2 = \frac{1}{2}\delta_{ij}\dot{\phi}_i\dot{\phi}_j + V \, , \\ \label{eq:hdot}
	& \dot{H} = - \frac{1}{2}\delta_{ij}\dot{\phi}_i\dot{\phi}_j  \, , \\ \label{eq:kl_go}
	& \ddot{\phi}_i + 3H \dot{\phi}_i + V_{,i} =0 \, .
	 \end{align}
The latter $1+\mathcal{N}$ equations form a non-linear system for the unknown functions $\phi(t)$ and $a(t)$ (or $H(t)$)  which is in general unsolvable for an arbitrary potential function. In the cosmic time parameterization the  ODE \eqref{eq:kl_go} becomes separable for a constant potential, which describes a de Sitter universe. 

In Ref.~\cite{Chimento:1998ju} the general solution for a two-field model consisting of one massless and one self-interacting field with an exponential potential, $V(\phi,\psi) = V_0 e^{p\phi}$, was first constructed. Using the conservation of canonical momentum the problem was shown to be equivalent to a single exponential potential and stiff matter. Equation \eqref{eq:hdot} becomes
\begin{equation}
\dot{H} = -{1 \over 2}\dot{\phi}^2 - {1 \over 2}{c_0 \over a^6} \, .
\end{equation}
and in combination with Eq.~\eqref{eq:fc} provides the solution for $a$. It is worth noticing that the same solution holds for an arbitrary number of  additional massless fields, whose influence can be absorbed in the definition of the constant $c_0$. In the presence of a non-zero gradient for the second field this method can not be applied (at least without serious modification).\footnote{It could be argued that an appropriate orthogonal rotation eliminates dependence of the potential on every field but one, arriving back to the case studied in \cite{Chimento:1998ju}. However, to map this solution with $p_j=0$ (for $j\neq1$) to the case of general gradients $p_j\neq0$ requires the explicit construction of the rotation $\mathcal{N} \cross \mathcal{N}$ matrix, a problem that is considerably more involved than just solving the system of equations without considering $p_j=0$, as we will show in the next section.}

In the following we adopt a different method that allows the derivation of the general solution of a product-exponential potential,  i.e.
\begin{equation} \label{eq:prod_exponential}
V= V_0 e^{\lambda_i \phi_i} \, ,
\end{equation}
for some constant numbers $\lambda_i$ and without assuming $\mathcal{N}-1$ of them to be zero.\footnote{Note that observables evaluated on the late-time solutions (as one typical does in all inflationary models) do not satisfy the constraints from the latest \textit{Planck} results, as they are identical to observables for single-field exponentials. These models, therefore, may be more relevant as dark energy candidates.}

\subsection{Integrability of the product-exponential potential}
For $\mathcal{N}$ canonically normalized fields switching to the e-folding number, defined from $N \equiv \ln a = \int H \ud t$, the Klein-Gordon system of Eq.~\eqref{eq:kl_go} can be written in first order form as
\begin{align} \label{eq:multi_fev}
	\phi_i' &= v_i \, ,
	\\ \label{eq:multi_vev}
	 v_i' &= - \left(3 - {1 \over 2}v_k v_k\right) \left(v_i + p_i \right)  \, ,
\end{align}
where we defined $p_i \equiv (\ln V)_{,i}$. The velocities in terms of the number of e-folds are related to the slow-roll parameter $\epsilon$ as
\begin{equation}
\epsilon = {1 \over 2} v_i v_i \, .
\end{equation}
In this formulation it becomes clear that the logarithmic derivatives of the potential (and not the gradients $V_{,i}$) determine the evolution of $\epsilon$. For the product-exponential potential of Eq.~\eqref{eq:prod_exponential} $p_i$'s are constant and Eqs.~\eqref{eq:multi_vev} decouple from Eqs.~\eqref{eq:multi_fev}, so one can study the reduced system. Even though Eqs.~\eqref{eq:multi_vev} remain coupled through the term ${1 \over 2}v_k v_k$, the system is fully integrable as there exist integrals of motion which enable us to further reduce the dynamical variables of the problem. These can be found by dividing Eqs.~\eqref{eq:multi_vev} for two fields e.g.~$i$ and $j$
\begin{equation} \label{eq:int_motion}
{\ud v_i \over \ud v_j} = {v_i + p_i \over v_j + p_j} \Rightarrow v_i = A_i (v_j + p_j) - p_i \, ,
\end{equation}
where we absorbed the initial conditions dependence in the definition of
\begin{equation}
A_i \equiv {v_{i,0} + p_i \over v_{j,0} + p_j} \, .
\end{equation}
For $\mathcal{N}=2$ and $p_i=0$ this integral of motion is identical to Eq.~(10) of Ref.~\cite{Chimento:1998ju}. Note that these integrals of motion make sense if the initial conditions for the two fields are not chosen exactly on the critical points of \eqref{eq:multi_vev}. If for instance $v_{k,0}=-p_k$ then the unique solution for that field is $v_{k}(N)=-p_k$ and hence it is not dynamical. Discarding this equation one then solves the reduced system of $\mathcal{N}-1$ equations.

\subsection{Asymptotic behaviour} \label{sec:critical_points}
Before deriving the solution we will study the asymptotic behaviour for $N\rightarrow \infty$. The system of Eqs.~\eqref{eq:multi_vev} with constant $p_i$ have the following set of critical points:
\begin{enumerate}
\item \label{crit2}  Power-law (or scaling) solution 
\begin{equation}
v_i = - p_i \, ,
\end{equation}
where the ratio of kinetic to potential energy becomes constant. This critical point exists for $p_i p_i <6$ and the critical value for $\epsilon$ is $1/2 p_i p_i$.

\item \label{crit1} Kinetic energy domination, i.e.
\begin{equation}
K\equiv {1 \over 2} \dot{\phi}_i \dot{\phi}_i \gg V \, ,
\end{equation} 
where the kinetic energy of the fields dominate their potential energy. The critical value for $\epsilon$ is 3, which defines an  $\mathcal{N}$-dimensional hypersphere and, hence, it is a collection of critical points. 

\end{enumerate}
The stability of these critical points trivially follows after linearization of the system around the critical points. For the scaling solution, translating the critical point to the origin by defining
\begin{equation}
x_i \equiv v_i+p_i \, ,
\end{equation}
then the linearized equation of motion reads
\begin{equation}
x_i' = - \left(3 - {1 \over 2} p_ip_i \right) x_i \, ,
\end{equation}
and hence we find negative eigenvalues when the condition $p_ip_i <6$ is satisfied. We observe that whenever this critical point exists it is also stable.

Regarding kinetic domination, the set of Eqs.~\eqref{eq:multi_vev} is not very useful because more than one realizations can give $\epsilon=3$. This also implies that critical points are not isolated and one has to apply more sophisticated methods to infer stability. For this reason, we will use the evolution equation for the slow-roll parameter which is found by contracting Eq.~\eqref{eq:multi_vev} with $v_i$ 
\begin{equation} \label{eq:depsilon_multi}
\epsilon ' = -(3-\epsilon)(2\epsilon + p_iv_i) \, .
\end{equation}
Defining a new variable 
\begin{equation}
x \equiv \epsilon-3 \, ,
\end{equation}
the linearized equation for $x$ becomes
\begin{equation} 
x' = x ( 6 + p_i v_i) \, ,
\end{equation}
and hence stable realizations of kinetic domination require
\begin{equation} \label{eq:condition_kinetic}
6 + p_i v_i<0  \, .
\end{equation}
We can view the second term as the dot product of the vector $p_i$ and the vector $r_i\equiv v_i/\sqrt{6}$, which denotes a point on the $\mathcal{N}$-dimensional hypersphere defined from $v_i v_i=6$. The condition \eqref{eq:condition_kinetic} can be written as
\begin{equation} \label{eq:kin_dom}
\left(\sqrt{6}+ \vec{p}\cdot \vec{r}\right)<0 \, ,
\end{equation}
and since $|\vec{r}|=1$, the previous inequality implies that 
\begin{equation}
p_i p_i >6 \, ,
\end{equation}
is necessary and sufficient condition for kinetic domination to be stable. The points of stable kinetic domination should satisfy the inequality \eqref{eq:kin_dom}.

Considering $N \rightarrow -N$, i.e.~moving backwards in time, the stability of these critical points is reversed. An important note is in order: because eigenvalues for stable solutions are strictly negative numbers, the flow towards the critical point is monotonic; this means that the velocities for individual fields would never cross their critical value when they approach it from higher or lower values. This observation will be used later in Sec.~\ref{subsec:analsol}.

\subsection{Derivation of the solution}
Having expressed all velocities in terms of e.g.~the reference field $v_1$ then in principle we can solve the equation of motion for $v_1$ and obtain the relation $v_1(N)$ and then using the integrals of motion we can find the expression for the rest fields. However, we find it more instructive to work with $\epsilon$ as it is a central variable in inflation and dark energy models and, as we will show later, it better highlights the behaviour of the solution. Therefore, in the following we will express all velocities in terms of the slow-roll parameter. Using the integrals of motion in the expression of the slow-roll parameter 
\begin{equation} \label{eq:eps_v1}
2\epsilon = v_i v_i = \sum_{i=1}^{\mathcal{N}} (A_i (v_1 + p_1) - p_i)^2   \, ,
\end{equation}
provides a quadratic equation for $v_1$ 
\begin{equation} \label{eq:quadr_eq}
\begin{aligned}
 \boldsymbol{A}^2  v_1^2 + 2 (\boldsymbol{A}^2 p_1 - A_i p_i) v_1 + \boldsymbol{A}^2 p_1^2- 2 p_1 A_i p_i  \\  
+ p^2 -2 \epsilon  =0 \, ,
\end{aligned}
\end{equation}
where we defined $\boldsymbol{A}^2\equiv A_i A_i$ and $p^2 \equiv p_ip_i$. The two solutions are
\begin{equation} \label{eq:v1root}
v_{1,\pm} = {A_i p_i   \over \boldsymbol{A}^2} - p_1 \pm {1 \over \boldsymbol{A}^2} \sqrt{ \boldsymbol{A}^2 (2\epsilon  - p^2)  +  (A_i p_i )^2 } \, .
\end{equation}
At this point we are unaware which root is the correct one and we need to compare this expression with another one; contracting Eq.~\eqref{eq:int_motion} with $A_i$ yields
\begin{equation}
v_1 = {A_i p_i  \over \boldsymbol{A}^2 }  - p_1  + {A_i v_i \over \boldsymbol{A}^2 }    \, ,
\end{equation}
and, hence, the last term of Eq.~\eqref{eq:v1root} is equal to
\begin{equation} \label{eq:plus_minus}
\pm \sqrt{ \boldsymbol{A}^2 (2\epsilon  - p^2)  +  (A_i p_i )^2 } = A_i v_i  \, .
\end{equation}
Thus, the sign of $A_iv_i$ determines which root is the correct one and the relation of $v_1$ in terms of $\epsilon$ is
\begin{equation} \label{eq:v1_sol_epsilon}
v_{1} (\epsilon) = {A_i p_i   \over \boldsymbol{A}^2} - p_1 + s {1 \over \boldsymbol{A}^2} \sqrt{ \boldsymbol{A}^2 (2\epsilon  - p^2)  +  (A_i p_i )^2 } \, ,
\end{equation}
where  
\begin{equation} \label{eq:sign}
s \equiv \sgn(A_i v_{i}) \, .
\end{equation}
Using this expression all velocities are written in terms of the slow-roll parameter 
and the differential equation of $\epsilon$
\begin{equation}\label{eq:deps}
\epsilon' = -(3 -\epsilon) \left[ 2\epsilon + p_i  A_i ( v_1(\epsilon)  +p_1) - p^2  \right] \, ,
\end{equation}
becomes separable. Note, however, that we are able to find the analytical solution of the inverse function $N(\epsilon)$
\begin{equation} \label{eq:nefmf} 
\begin{aligned}
 N(\epsilon) (6-p^2)   =& - 2 \ln \left( {A_i p_i + s \sqrt{ (A_i p_i)^2 + \boldsymbol{A}^2 ( 2\epsilon - p^2 ) } \over A_i p_i + s \sqrt{ (A_i p_i)^2 + \boldsymbol{A}^2 ( 2\epsilon_0 -p^2 )  }}  \right)  + \ln \left( {3 - \epsilon \over 3 - \epsilon_0 }\right)  +{s\cdot 2 A_i p_i \over \sqrt{(A_i p_i)^2 + \boldsymbol{A}^2 (6- p^2)} } \cdot \\ 
& \left[  \atanh \left( {\sqrt{ (A_i p_i)^2 + \boldsymbol{A}^2 ( 2\epsilon -p^2 ) } \over \sqrt{(A_i p_i)^2 + \boldsymbol{A}^2 (6- p^2)}}  \right)  - \atanh \left( {\sqrt{ (A_i p_i)^2 + \boldsymbol{A}^2 ( 2\epsilon_0 -p^2 )  } \over \sqrt{(A_i p_i)^2 + \boldsymbol{A}^2 (6- p^2)}}  \right) \right]  \, . 
\end{aligned}
\end{equation}
When the initial velocities of every field, except for $v_1$, take their critical value $v_{j,0}=-p_j$, then these fields are non-dynamical giving $A_i=0$ and one obtains the single-field solution. 

The field displacement as a function of the slow-roll parameter can be obtained by Eq.~\eqref{eq:deps} after a time redefinition $N \rightarrow \phi_j$
\begin{equation}
{\ud \epsilon \over \ud \phi_{(j)}} v_{(j)} = - (3 -\epsilon) \left[ 2\epsilon + p_i  A_i ( v_1(\epsilon)  +p_1) - p^2  \right] \, ,
\end{equation}
where no sum in j is implied. This equation is separable and as previously the inverse function $\phi_j(\epsilon)$ can be integrated with solution

\begin{equation}
\begin{aligned}  
 \Delta \phi_j (\epsilon) (6-p^2)  = & 2 p_j  \ln \left( {A_i p_i + s \sqrt{ (A_i p_i)^2 + \boldsymbol{A}^2 ( 2\epsilon - p^2 ) } \over A_i p_i + s \sqrt{ (A_i p_i)^2 + \boldsymbol{A}^2 ( 2\epsilon_0 -p^2 )  }}  \right) - p_j \ln \left( {3 - \epsilon \over 3 - \epsilon_0 }\right)  - {s\cdot 2 \left( A_j (6-p^2) + (A_i p_i) p_j   \right)\over \sqrt{(A_i p_i)^2 + \boldsymbol{A}^2 (6- p^2)} }  \\    & \left[  \atanh \left( {\sqrt{ (A_i p_i)^2 + \boldsymbol{A}^2 ( 2\epsilon - p^2 ) } \over \sqrt{(A_i p_i)^2 + \boldsymbol{A}^2 (6- p^2)}}  \right)   - \atanh \left( {\sqrt{ (A_i p_i)^2 + \boldsymbol{A}^2 ( 2\epsilon_0 -p^2 )  } \over \sqrt{(A_i p_i)^2 + \boldsymbol{A}^2 (6- p^2)}}  \right) \right]  \, . \label{eq:fi_mf}
\end{aligned} 
\end{equation}
With the expressions \eqref{eq:nefmf} and \eqref{eq:fi_mf} we have fully determined the solution $\{ a,\phi_j\}$ in terms of $\epsilon$.

\subsection{Analysis of the solution} \label{subsec:analsol}

The solution \eqref{eq:nefmf} elegantly captures the past or future asymptotic behaviour of the Klein-Gordon equations. Here one has to distinguish between two cases, depending on the sign of $s$ relative to $A_i p_i$.

\begin{enumerate} 

\item  \label{case_A} When $s\cdot(A_i p_i)>0$, the second and fourth terms diverge to minus infinity for $\epsilon \rightarrow 3$, proving that  kinetic energy domination is the past attractor.

\item \label{case_B} When $s \cdot (A_i p_i)<0$, there are two more subcases:

\begin{enumerate}

\item $p^2<6$. The first term diverges as $\epsilon \rightarrow 1/2 p^2$ and the scaling solution is the future attractor.

\item $p^2>6$. The scaling solution does not exist and kinetic domination is the future attractor.

\end{enumerate} 

\end{enumerate} 
The previous analysis indicates that if initial conditions belong to case \ref{case_A} then $s$ should change to $-s$ in order for the system to reach one of the critical points described in case \ref{case_B}.

Now we will briefly explain how one can use the full expression \eqref{eq:nefmf}, while \eqref{eq:fi_mf} can be treated similarly. We follow the following steps:
\begin{enumerate}
\item For given initial conditions $\{ v_{i,0}\} $ we calculate $s$ and the slow-roll parameter $\epsilon_0$. Depending on the sign of $s$ we choose the appropriate branch in Eq.~\eqref{eq:nefmf}. For the specified initial conditions the solution will then be given as a function of $\epsilon_0$ and $s$: $ N= N(s,\epsilon_0,\epsilon)$.
\item We calculate the value of $\epsilon$ for which $s$ changes. This would happen for a specific value of $\epsilon = \epsilon_*$, found by setting Eq.~\eqref{eq:plus_minus} to zero:
\begin{equation}
\epsilon_*  = {1 \over 2}p^2  - {(A_i p_i )^2 \over 2 \boldsymbol{A}^2} \, .
\end{equation}
Although this value always exists (due to the Schwarz inequality $(A_i p_i)^2 \leq A^2 p^2$), it may not be relevant for the assumed set of initial conditions. This is because the system may relax at one of its critical points before crossing this value.
\item  Next, we compare the sign of $A_ip_i$ with $s$. If $s =\sgn{(A_ip_i)}$ then we move to step \ref{step_pos} and otherwise to step \ref{step_neg}.
\item  \label{step_pos} For this set of initial conditions the full solution will consist of both branches, namely 
\begin{align}
&N(\epsilon,\epsilon_*,s)\, , ~~\text{for}~~\epsilon \in (\epsilon_*,3) \, , \\
&N(\epsilon,\epsilon_*,-s)\, , ~~\text{for}~~\epsilon \in (\epsilon_*,\epsilon_{\rm cr}) \, .
\end{align}
where $\epsilon_{\rm cr}$ corresponds to the logarithmic divergence of the solution. Of course, we know from the asymptotic analysis that the divergence corresponds to one of the two types of critical points, that is either kinetic domination or the scaling solution.
\item \label{step_neg} In this case we have to compare $\epsilon_*$ with $\epsilon_{\rm cr}$, that corresponds to the logarithmic divergence of the solution. If $\epsilon_*<\epsilon_{\rm cr}<\epsilon_0$ then $\epsilon$ would never reach 
$\epsilon_*$ and the solution is 
\begin{align} \label{eq:negative1}
&N(\epsilon,\epsilon_{0},s)\, , ~~\text{for}~~\epsilon \in (\epsilon_{\rm cr},3) \, .
\end{align}
In any other case the solution is 
\begin{align}
&N(\epsilon,\epsilon_*,s)\, , ~~\text{for}~~\epsilon \in (\epsilon_*,\epsilon_{\rm cr}) \, , \\
&N(\epsilon,\epsilon_*,-s)\, , ~~\text{for}~~\epsilon \in (\epsilon_*,3) \, .
\end{align}

\end{enumerate}

\begin{figure}[t!]
	\includegraphics[scale=0.7]{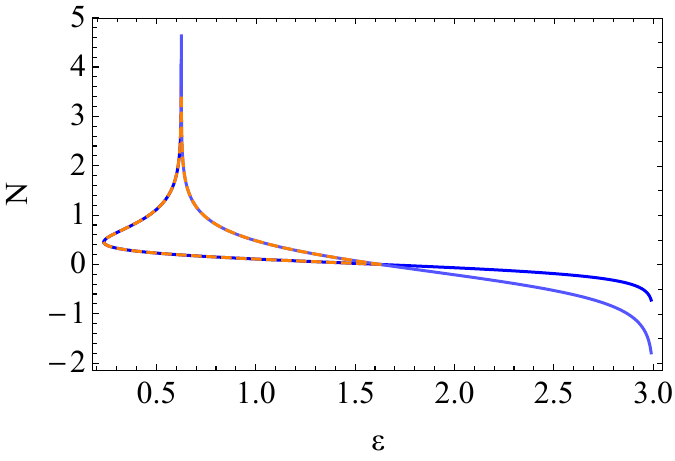}
	\caption{Evolution for $\mathcal{N}=50$ fields using the analytical solution (solid lines) versus the solution obtained with numerical integration (orange dashed lines). The solid lines extrapolate up to kinetic domination showing that is the past attractor.}
	\label{fig:50fields}
\end{figure}

In Fig.~\ref{fig:50fields} we plot the analytical solution \eqref{eq:nefmf} for 50 fields, along with the numerical result, for a model with $p_1=1,~p_2=0.5,~p_j=0.05$ $(j=3,\cdots,50)$ for two different sets of initial conditions. In the first one  we consider $v_1=-1.5,~v_2=-1,~v_j=0.01$ yielding $\epsilon_0=1.685,~\epsilon_*=0.552,~s=-1$ and $\sgn(A_i p_i) =1$ and, thus, we are in step \ref{step_neg}. Comparing $\epsilon_*$ with the asymptotic value $1/2p_ip_i = 0.685$ we observe that $\epsilon_*<\epsilon_{\rm cr}<\epsilon_0$ and the analytical solution is given by \eqref{eq:negative1}. This is depicted with the pale blue line in Fig.~\ref{fig:50fields}; Moreover, we observe that the solution extrapolates from kinetic domination to the scaling solution located at $\epsilon=0.685$, but it never crosses $\epsilon_*$. For our second set of initial conditions we consider $v_1=1.5,~v_2=1,~v_j=0.01$ and so we find $\epsilon_0=1.685,~\epsilon_*=0.007,~s=1$ and $\sgn(A_i p_i) =1$ and we are in step \ref{step_pos}. The solution consists of two branches, one with $s=1$ defined in the interval $\epsilon \in (0.007,3)$ and another with $s=-1$ defined in the interval $\epsilon \in (0.007,0.685)$. Again, the solution extrapolates from kinetic domination to the scaling solution crossing $\epsilon_*$.

Finally, we provide some physical insight on the terms $A_i v_i$ and $A_i p_i$. Their physical significance becomes more apparent if we examine the single-field case where $A_i=1$ and the two terms become simply $v_1$ and $p_1$. When e.g~$0<p_1<\sqrt{6}$ then the asymptotic value for $v_1$ would be $v_1=-p_1$. If the initial velocity $v_1$ is positive then it first has to decrease to zero and then decrease further until it reaches the value $-p_1$. Eq.~\eqref{eq:v1_sol_epsilon} becomes simply 
\begin{equation} \label{}
v_{1} (\epsilon) =  s  \sqrt{ 2\epsilon  } \, ,
\end{equation}
and  when s becomes zero we need to switch to the other branch of the solution. For more fields the distinction between the two case is more complicated because all velocities are mixed through the $A_iv_i$ terms.

\section{From kinetic domination to accelerating expansion}\label{subsec:ab}
\subsection{How many e-folds before inflation?}

For the product-exponential potential the existence of an analytical solution allows us to calculate the number of e-folds the system spends before it approaches the asymptotic solution. For accelerating solutions specifically, $p_i$ should be chosen such that $1/2p_ip_i<1$. If the system starts at a state with $\epsilon_0>1$, then the solution \eqref{eq:nefmf} provides the necessary number of e-folds before inflation begins. 

For an arbitrary potential, one can ask a similar question; assuming that there exists an ``asymptotic'' state (in practice the slow-roll solution) and that the system begins with initial conditions close to kinetic domination, then how many e-folds are necessary before inflation starts? The exact duration for the passage from a state with $\epsilon>1$ to a state with $\epsilon<1$  requires the full analytical solution, which seems an impossible task. To make some progress, we can construct an estimate based on the previously derived solution \eqref{eq:nefmf}. This relies on the observation that for small field displacements any continuous function $p_i$ can be approximated by its constant part (the first term in the Taylor series) and this gives a first estimate on the total number of e-folds. However, since the value of $p_i$ changes we do not know a priori which value to choose. In the following we will show that in general the minimum and maximum values of $p_i$ give the minimum and maximum number of e-folds respectively. To prove this we will first derive some properties of the solutions for product-exponential potentials.

\begin{figure}[t!]
	\centering
	\includegraphics[scale=0.6]{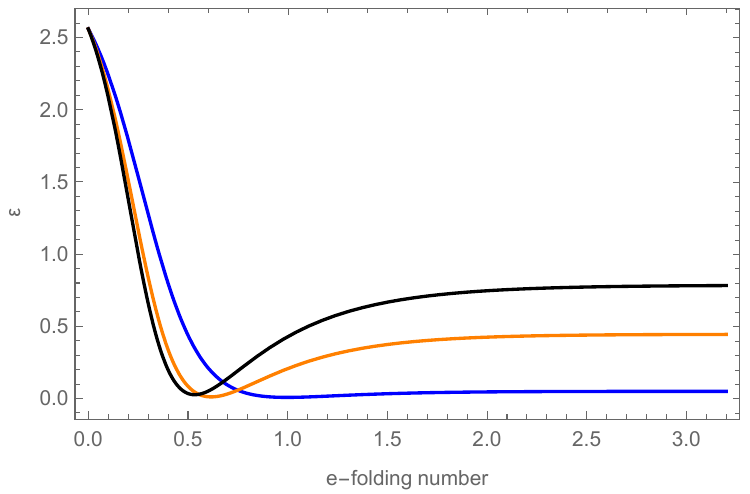}
	\includegraphics[scale=0.6]{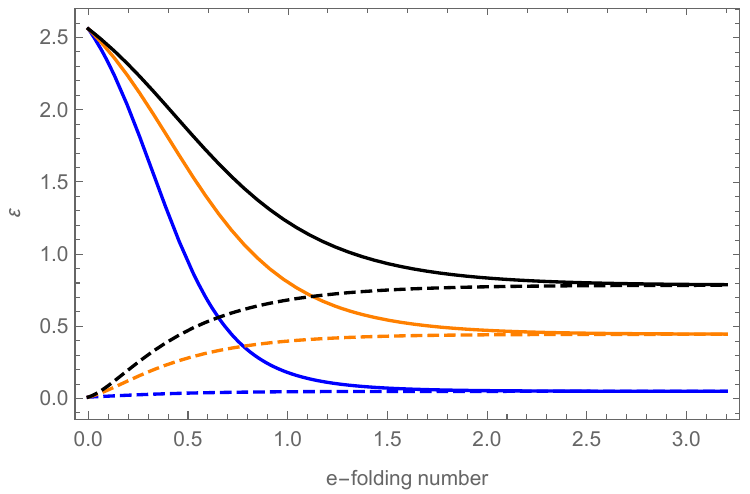}
	\caption{Evolution of 3 product-exponential potentials with $(p_1,p_2)$ $=(0.1,0.3)$, $(0.5,0.8)$, $ (0.6,1.1)$ (blue, orange and black accordingly). Upper graph: initial velocities are $v_{i,0}=1.6$. Lower graph: initial velocities are negative with $v_{i,0}=-1.6$ (solid lines) or $v_{i,0}=-0.1$ (dashed lines) indicating supercritical or subcritical initial values.}
	\label{exp4}
\end{figure}

In  Fig.~\ref{exp4} we plot the solution for three different exponential potentials ($V_{p}\propto e^{p_i \phi_i}$ and $V_{\lambda}\propto e^{\lambda_i \phi_i}$) satisfying $0<p_i<\lambda_i$ and for positive (upper plot) or negative (lower plot) initial velocities that are close to kinetic domination. In the first case, $v_{(i)} p_{(i)} , v_{(i)} \lambda_{(i)}> 0$ holds for each $i$ (with no sum implied), while in the second one $v_{(i)} p_{(i)} , v_{(i)} \lambda_{(i)}<0 $. In the upper plot, $\epsilon$ first decreases until the value $\epsilon_*$ and then it increases again until it reaches its asymptotic value. During the decreasing period we observe that steeper gradients lead to faster decrease, whereas this behaviour is reversed once $\epsilon$ starts to increase. This makes sense physically because in the first case the Hubble friction cooperates with the gradient term (until the velocities change sign) and hence $\epsilon$ decreases faster for steeper gradients; similarly, when the velocities increase steeper gradients lead to larger terminal velocities and hence the system needs more time to reach those values. However, in both plots the system reaches its asymptotic state slower for steeper gradients. 

In certain cases we can understand the previous behaviour analytically. For two different product-exponential potentials with $0<|p_i| < |\lambda_i|$ we obtain $0<|v_i p_i| < |v_i \lambda_i|$. When each velocity $v_i$ has sign opposite to $p_i,\lambda_i$ then
\begin{equation}
|p_i| < |\lambda_i| \Rightarrow v_i p_i > v_i \lambda_i \, ,
\end{equation}
and we obtain the following inequality
\begin{equation} 
(3-\epsilon)(2\epsilon  + p_i v_i) > (3-\epsilon) (2\epsilon  + \lambda_i v_i) \, .
\end{equation} 
Using equation \eqref{eq:depsilon_multi} it is transformed into a differential inequality 
\begin{equation} \label{eq:sq}
{\ud \epsilon \over \ud N} < {\ud \tilde{\epsilon} \over \ud N}  \Rightarrow \epsilon(N)< \tilde{\epsilon}(N) \, ,
\end{equation}  
where we solved the differential equations with equal initial conditions $\epsilon(N_0) = \tilde{\epsilon}(N_0)$. In the opposite case where velocities have all the same sign with $p_i,\lambda_i$ then the previous inequality is reversed. In order for this inequality to be satisfied $\epsilon$ should adjust the rate of change according to the magnitude of the gradients. This is exactly what is depicted e.g.~in the lower plot of Fig.~\ref{exp4}; we observe that for the three sets of exponents $\epsilon(N)$ follows the same inequalities and hence, schematically we have black > orange >blue.

A different way to view the previous is that for a given set of $p_i$ we can find two different sets of exponents, e.g.~$\lambda_i$ and $\kappa_i$, whose solutions bound the solution for $p_i$ at all times.  In the derivation of these relations the strict inequality between the different pairs of exponents is crucial; the condition on the norm of the gradient vector $|\kappa|^2 <|p|^2 < |\lambda|^2$ is not sufficient to provide a bound on $\epsilon$ at all times (see Fig.~\ref{exp4}). The inequalities should instead be satisfied for each individual component.  Moreover, we required that each velocity should have either the same sign as the corresponding gradient or opposite sign; if this is not satisfied  this method may fail to provide definite inequalities for $\epsilon$. This property of the solutions can provide some conclusions about the evolution of more general potentials.  For field-dependent $p_i(\phi_j)$, we can estimate the number of e-folds by considering the minimum and maximum values for $p_i(\phi_j)$ in some field interval $[\phi_{j,\rm min}, \phi_{j,\rm max}]$. Using  these values we calculate $ N_{\rm  min}$ and  $N_{\rm max}$ that correspond to the the solutions for product exponential potentials ($e^{p_{i,\rm  min}\phi_i}$ and $e^{p_{i,\rm max}\phi_i}$ respectively) and then estimate the actual transitional time in the interval $N\in (N_{\rm min},N_{\rm max})$.
\begin{figure}[t!]
	\centering
	\includegraphics[scale=0.6]{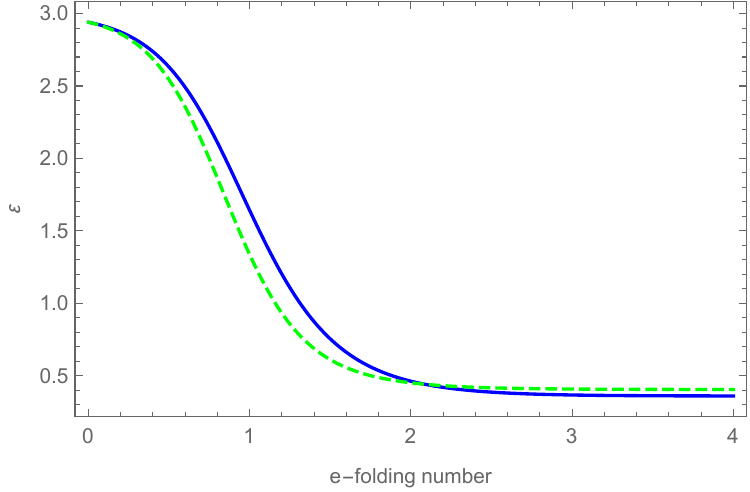}
	\caption{Evolution of 2 product-exponential potentials with $(p_1,p_2)= (0.6,0.6)$ and $(\lambda_1,\lambda_2) = (0.01,0.9) $ (blue and green dashed accordingly) for negative supercritical velocities with the same $\epsilon_0$. Even though $|p|^2<|\lambda|^2 $ the two lines cross.}
	\label{exp44}
\end{figure}
\begin{figure}[t]
	\centering
	\includegraphics[scale=0.6]{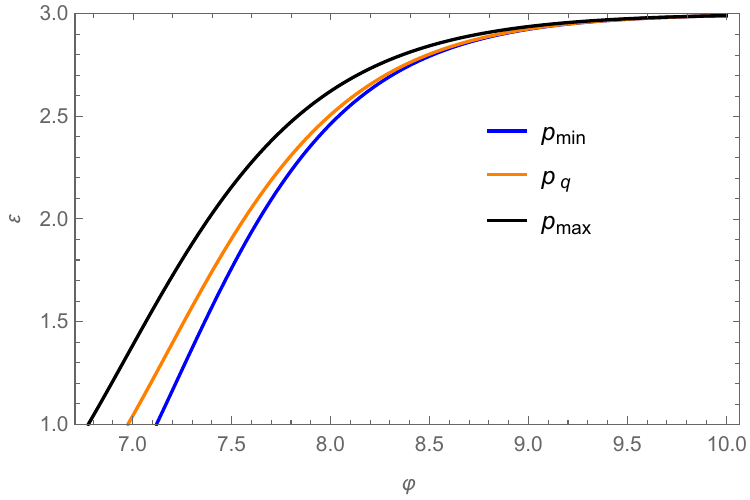}
	\caption{Estimation of $\epsilon$ for a quartic potential using two exponential potentials with $p_{min}=4/\phi_0$ and $p_{max}=4/\phi_*$ for negative velocities, $\epsilon_0=2.99$  and $\phi_0=10 M_{pl}$.}
	\label{fig:estimate}
\end{figure}

As an illustrative example we will study quartic inflation $V\sim \phi^4/4$. We set the initial conditions to $\phi_0=10$ and $\epsilon_0=2.99$. The field flows to the origin and in the interval $(0,\phi_0)$  the gradient $p=4/\phi$ is a monotonically decreasing function; therefore, the minimum value for $p$ would be given by $p(\phi_0)$, whereas the maximum value corresponds to the value of $\phi$ which gives $\epsilon(\phi_*)=1$. The latter value is unknown, as we do not know the full solution; in the following we will demonstrate how to estimate it. The value $\phi_*$ would be the  displacement one would find using the analytical solution \eqref{eq:fi_mf} if the value $p(\phi_*)$ was used for the gradient term
\begin{equation}
\phi_* = \phi_0 + \Delta\phi(\epsilon=1,\epsilon_0,p(\phi_*)) \, .
\end{equation}
Note that the previous is a transcendental equation for $\phi_*$. Once this value is known, then $p_{\rm max}=p(\phi_*)$ can be used in \eqref{eq:nefmf} to find the maximum number of e-folds. The precision of this estimate is portrayed in Fig.~\ref{fig:estimate} and we can predict a priori the field displacement between 2.9-3.2 in units of $M_{\rm pl}$, whereas the usual estimation resulting from considering kinetic domination with $p=0$ gives $\Delta \phi = 2.35 ~M_{\rm pl}$. This difference, although small, can potentially be important in the light of different conjectures which severely restrict the allowed superplankian field displacement, as well as the minimum value of $p$ \cite{Obied:2018sgi,Agrawal:2018own}. Conversely, at a given point $\phi_0$ one can constrain $\epsilon_0$ by the requirement of 50-60 e-folds of inflation and therefore estimate the ``basin of attraction'' of the initial conditions that provide sufficient inflation.

\subsection{The gradient-flow approximation}

For potentials with slowly varying gradients (in some field interval) the two estimated values for $\epsilon$ will be very close. This implies that the system behaves closely to a product exponential with field-dependent exponents, which, nevertheless, satisfy $p_i'\ll 1$. Moreover, the scaling relations for the velocities $v_i \approx -p_i$ imply that $\epsilon$ varies slowly 
\begin{equation} \label{mfsr}
\left({1 \over 2 }p_ip_i \right)' \approx  - { V_{,k} V_{,i} V_{,ik} \over V^2} + (2 \epsilon_V)^2 \, ,
\end{equation}
and the two terms should be small. In the latter equation we recognise the conditions for the validity of the gradient-flow approximation \cite{Lyth:2009zz,Yang:2012bs,GrootNibbelink:2000vx,GrootNibbelink:2001qt,Peterson:2010np,Peterson:2011ytETAL}. Therefore, models that are approximated well by gradient flow lie at the intersection between de Sitter and scaling solutions. The reason why these models asymptote to a state with almost vanishing accelerations can be explained by the fact that they are small deformations of de Sitter-scaling models, which have identically zero accelerations; since they are not exactly product-exponentials, during their evolution they move adiabatically between asymptotic solutions of exponentials.\footnote{This observation has also been used in a subsequent paper which focused on the regime of large turn rate and non-trivial field-space geometry \cite{Christodoulidis:2019jsx}.} In this way, one can deduce the global convergence towards the slow-roll solution which can not be done using linearised analysis. As a final remark, the scaling relations fix the fields' velocities, through $v_i \approx -p_i$, but not their position. Therefore, under this type of approximation observables inherit an explicit dependence on (a subset of) initial conditions.

\section{Generalizations} \label{sec:generalization}
In addition to the scalar fields we will consider a non-interacting fluid with an equation of state $P = w \rho$. The Klein-Gordon equations remain unchanged, while Eqs.~\eqref{eq:hdot} and \eqref{eq:kl_go} are modified 
\begin{align}
3H^2 &= \frac{1}{2} \dot{\phi}_i\dot{\phi}_i + V + \rho \, . \\
 \dot{H} &= - \frac{1}{2} \dot{\phi}_i\dot{\phi}_i  -\frac{1}{2} (1 + w) \rho \, ,
\end{align}
and we obtain an additional equation, the conservation equation for the fluid,
\begin{equation}
\dot{\rho} + 3H (1 + w) \rho =0 \, .
\end{equation}
The slow-roll parameter is now given by
\begin{equation} \label{eq:slow_roll_fluid}
\epsilon = \frac{1}{2} v_i v_i +{1 \over 2}(w+1) {\rho \over H^2} \, .
\end{equation}
Introducing a new variable $z\equiv \rho/H^2$ and using 
\begin{equation}
{V \over H^2} = 3 - \epsilon + {1 \over 2} (w-1) z \, ,
\end{equation}
the system of evolution equations can be written as
\begin{align}
& v_i ' +(3-\epsilon)(v_i + p_i) + {1 \over 2}(w-1)z p_i =0 \, , \\ \label{eq:fluid_w1}
& z' + \left( 3+3w -2\epsilon \right) z =0 \, .
\end{align}
In this form one can immediately read off asymptotic solutions, generalizing the solutions for one field (see e.g.~\cite{Copeland:1997et})
\begin{enumerate}
\item scalar kinetic domination: $z=0$, $\epsilon =3$. There exists an infinite number of field realizations describing kinetic domination, namely all points on the hypersphere satisfying $v_iv_i=6$. 

\item scalar potential domination: $z=0$, $v_i = -p_i$ with $\epsilon =1/2 p_i p_i$.

\item fluid domination: $v_i = 0$, $z=3$ with $\epsilon =3/2(1+w)$.

\item scaling solution: fields and fluid are non-zero with $\epsilon =3/2(1+w)$ (the expressions for $v_i$ and $z$ are more complicated but can be found by solving quadratic equations).
\end{enumerate}
In the case of a fluid with $w=1$~\footnote{This fluid resembles the effective equation of state of a scalar field in the kinetic domination regime or in the absence of a potential ($V=0$). However, the problem is not equivalent to a $(\mathcal{N}+1)$-field model, where the extra field is massless, due to the extra factor of 2 in Eq.~\eqref{eq:fluid_w1} (see also \cite{Faraoni:2012hn}).} equations simplify to
\begin{align}
& v_i ' +(3-\epsilon)(v_i + p_i)  =0 \, , \\
& z' + \left(3 - \epsilon \right) 2z =0 \, ,
\end{align}
and, hence, one obtains another integral of motion which can be used to express the fluid in terms of a reference field
\begin{equation}
z = {z_0  \over \left(v_{1,0 } + p \right)^2 }  \left( v_1 + p  \right)^2  \equiv {B \over 2}  \left( v_1 + p  \right)^2 \, .
\end{equation}
Plugging back into the slow-roll parameter \eqref{eq:slow_roll_fluid} we obtain
\begin{equation}
2 \epsilon = \sum_{i=1}^{\mathcal{N}} (A_i (v_1 + p_1) - p_i)^2  + B (v_1 + p_1)^2 \, ,
\end{equation}
which has the same form as Eq.~\eqref{eq:quadr_eq} after the substitution $A^2 \rightarrow \tilde{A}^2 = A^2 +B $.
Moreover, for $w=1$ Eq.~\eqref{eq:depsilon_multi} remains unchanged and the general solution in terms of $\epsilon$ has exactly the same form (Eqs.~\eqref{eq:nefmf} and \eqref{eq:fi_mf}) modulo the substitution $A^2 \rightarrow \tilde{A}^2$.

\section{Summary and discussion} \label{sec:concl}

Scalar fields in cosmology have been extensively studied over the last decades mainly due to their applications in both the early and the late universe. To date, only a limited number of exact solutions exists in the literature. In this paper we derived the general solution for an interacting product-exponential potential, that has not been considered in earlier works,  in terms of the slow-roll parameter $\epsilon$. In the latter form it better highlights properties of its critical points.

We argued how the solution becomes relevant in the derivation of estimates for the transitional time to inflation for models that follow the potential gradient flow. For the non-linear evolution towards the slow-roll solution, i.e.~starting with arbitrarily large deformations, one needs the full solution. Since there is no general prescription on how to solve the Klein-Gordon equation, one can alternatively derive estimates for the time of this transition using the solution of exponential potentials. This method generalizes the existing estimates, as it does not assume kinetic domination or negligible potential gradients $K\gg V,V_{,\phi}$ (that was considered in previous works, e.g.~\cite{Goldwirth:1991rj}). We showed that the actual transitional time will lie between two bounds that depend on magnitude of the potential gradient $(\ln V)_{,\phi}$. Besides providing more accurate results the method can be applied in multi-field models where linearized stability becomes obscure due to the coupling of fields' perturbations. 

It is worth noticing that the general solutions presented in this work can also describe multi-field domain walls. The correspondence between cosmological and domain wall solutions can be found e.g.~in the Refs.~\cite{Skenderis:2006rr,Skenderis:2007sm,McFadden:2009fg,Garriga:2014fda}. The beta function is the analogue of the slow-roll parameter $\epsilon$ and hence the RG flow of multi-field domain walls with product exponential potentials can be found using the solutions \eqref{eq:nefmf} and \eqref{eq:fi_mf} and the dictionary between cosmology and domain walls.

Finally, we considered the case of an extra non-interacting fluid. When the fluid obeys an equation of state $w=1$ (stiff matter) another integral of motion exists which makes the problem again fully integrable with almost the same solution as before (more precisely with a minor subsitution).\footnote{This solution, along with the solutions presented in Refs.~\cite{Chimento:1998ju,Paliathanasis:2018vru}, are perhaps the only examples in the literature of general solutions for an arbitrary number of fields.} It would be interesting to investigate how this solution can be generalized for problems with non-trivial field manifolds.

\section*{Acknowledgments}{It is a pleasure to thank Diederik Roest, Vincent Vennin, Vasil Rokaj and Vaios Ziogas for
helpful discussions and valuable comments on the first versions of the draft. I would also like to thank the anonymous referee for suggestions which helped me in improving the clarity of this work. I acknowledge financial support
from the Dutch Organisation for Scientific Research (NWO).}

\end{document}